%% file: RuO2_CoF2_IFE.tex
\documentclass[
    aps,
    prb,
    twocolumn,
    floatfix,nofootinbib,
    superscriptaddress
]{revtex4-2}

\usepackage{graphicx}
\usepackage{subfigure}
\usepackage[utf8]{inputenc}
\usepackage{times}
\usepackage[breaklinks,colorlinks=true]{hyperref}
\hypersetup{linkcolor=red,citecolor=green,urlcolor=blue}
\usepackage{color}
\usepackage{siunitx}
\usepackage{booktabs}
\usepackage[version=4]{mhchem}
\usepackage{ulem}
\usepackage{threeparttable}

\usepackage{amssymb} 
\usepackage{amsmath}
\usepackage{amsthm}
\usepackage{braket}
\usepackage{MnSymbol}
\usepackage{stmaryrd}
\usepackage{mathtools}

\input{commands}
\usepackage{tikz}
\usepackage{tikz-cd}
\usetikzlibrary{calc}
\usetikzlibrary{decorations.markings}
\usepackage{xcolor}
\usepackage{graphicx}

\makeatletter
\renewcommand\@biblabel[1]{#1.}
\makeatother
\begin{document}

\setcounter{secnumdepth}{2} 

\title{Spin and Orbital Magnetism by Light in Rutile Altermagnets}

\author{T. Adamantopoulos}
    \thanks{t.adamantopoulos@fz-juelich.de}
    \affiliation{\pgi}
    \affiliation{\aachen}

\author{M. Merte}
    \affiliation{\pgi}
    \affiliation{\aachen}
    \affiliation{\mainz}

    \author{F. Freimuth}
    \affiliation{\mainz}

\author{D. Go}
    \affiliation{\mainz}

\author{M. Le\v{z}ai\'c}
    \affiliation{\pgi}
    
\author{W. Feng}
    \affiliation{\moe}
    \affiliation{\beijing}
    
\author{Y. Yao}
    \affiliation{\moe}
    \affiliation{\beijing}
    
\author{J. Sinova}
    \affiliation{\mainz}
    \affiliation{\prague}

\author{L. \v{S}mejkal}
    \affiliation{\mainz}
    \affiliation{\prague}
    
\author{S. Bl\"ugel}
    \affiliation{\pgi}
    
\author{Y. Mokrousov}
    \affiliation{\pgi}
    \affiliation{\mainz}

\date{\today}

\begin{abstract}
While the understanding of altermagnetism is still at a very early stage, it is expected to play a role in various fields of condensed matter research, for example spintronics, caloritronics and superconductivity. In the field of optical magnetism, it is still unclear to which extent altermagnets as a class can exhibit a distinct behavior. Here we choose RuO$_2$, a prototype metallic altermagnet with a giant spin splitting, and CoF$_2$, an experimentally known insulating altermagnet, to study the light-induced magnetism in rutile altermagnets from first-principles. We demonstrate that in the non-relativisic limit the allowed sublattice-resolved orbital response exhibits symmetries, imposed by altermagnetism, which lead to a drastic canting of light-induced moments.  On the other hand, we find that inclusion of spin-orbit interaction  enhances the overall effect drastically, introduces a significant anisotropy with respect to the light polarization and strongly suppresses the canting of induced moments. Remarkably, we observe that the  moments induced by linearly-polarized laser pulses in light altermagnets can even exceed in magnitude those predicted for  heavy ferromagnets exposed to circularly polarized light.
By resorting to microscopic tools we interpret our results in terms of the altermagnetic spin splittings and of their reciprocal space distribution. Based on our findings, we speculate that optical excitations may provide a unique tool to switch and probe the magnetic state of rutile altermagnets.

\end{abstract}

\maketitle






\date{\today}


%
%
%
%
%
%
%
%
%
%
%
%
%
%
%


{\bf Introduction} 

Recently, altermagnets emerged as a new class of magnetic materials, which combine staggered collinear spin moments in real space, with an alternating spin splitting in the reciprocal space, leading to a perfect compensation of moments as a result of a combined action of time-reversal ($\mathcal{T}$), or, generally spin rotation symmetries, and crystal symmetries~\cite{Smejkal_PRX_2022a, Smejkal_PRX_2022b}. Due to the inherent spin-splitting of the states of d-, g-, or i-wave form, altermagnets are anticipated to host numerous exotic physical phenomena $-$ a fact that renders them as an extremely attractive platform in the context of spintronics. So far, several works revealed the existence of the anomalous Hall effect in a series of altermagnetic materials, e.g. RuO$_2$~\cite{Smejkal_2020, Feng_2022}, MnTe~\cite{Betancourt_2023}, SrRuO$_3$~\cite{Samanta_2020}, $\kappa$-type organic conductors~\cite{Naka_2020}, and perovskites~\cite{Naka_2022}, as well as generation of strong spin-polarized currents~\cite{Naka_2019, Bose_2022}, spin-splitter torque~\cite{Hernandez_2021, Bai_2022, Karube_2022}, spin-pumping~\cite{Sun_2023}, and crystal thermal transport~\cite{Zhou_2024}. The prospect of altermagnets in the field of optical magnetism has slowly started gaining interest and some initial experimental works studied their behavior upon laser excitation~\cite{Mashkovich_2021, Formisano_2022, Qiu_2023, Liu_2023}. Theoretically, on the other hand, very little is known about the interaction of altermagnets with light.

\begin{figure*}[ht!]
\begin{center}
\rotatebox{0}{\includegraphics [width=0.95\linewidth]{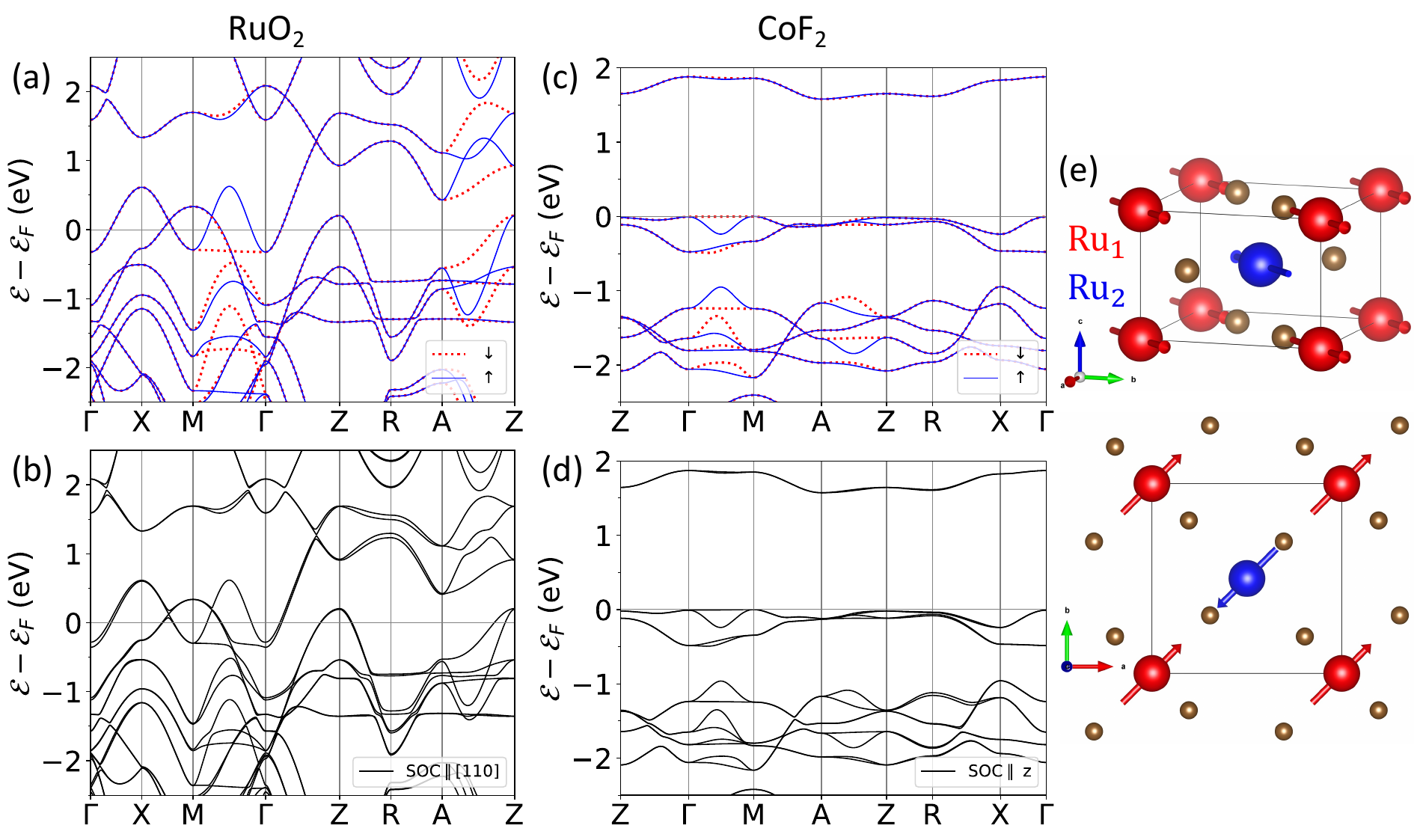}}
\end{center}
\caption{{\bf Electronic structure of altermagnetic $\mathbf{RuO_2}$ and $\mathbf{CoF_2}$}. (a-b) Bandstructures of  non-relativistic (a) and relativistic RuO$_2$  with $\mathbf{N}$ along  [110] (b).  (c-d) Bandstructures of non-relativistic (c) and relativistic CoF$_2$ with $\mathbf{N}$ along the $z$ axis (d). (e) Rutile crystal structure of RuO$_2$ with staggered magnetization along the [110] direction. The red and blue spheres indicate the Ru atoms in the opposite magnetic sublattices with the respective arrows depicting the magnetic moments. The golden spheres indicate the O atoms. The crystal structure of CoF$_2$ is identical to that of RuO$_2$.}
\label{Fig1}
\end{figure*}

Among the altermagnetic materials, tetragonal rutile RuO$_2$ with a $P4_2/mnm$ space group [see Fig.~\ref{Fig1}(g)] stands as one of the most explored. The theoretically predicted large spin-splitting of $1.3$\,eV~\cite{Smejkal_PRX_2022b,Smejkal_2020, Fedchenko_2024} labels RuO$_2$ as an ideal platform for the observation of emergent altermagnetic phenomena. Despite traditionally being considered a Pauli paramagnet, recently a room-temperature collinear antiparallel magnetic ordering was discovered in both RuO$_2$ bulk crystals and thin films~\cite{Berlijn_2017, Zhu_2019}, however, the reported small local magnetization value of $0.05\,\mathrm{\mu_B}$ still remains an open controversial issue~\cite{Smolyanyuk_2023}. Among rutile altermagnets another popular representative is CoF$_2$, which belongs to a family of well-known insulating transition-metal difluorides~\cite{Stout_1954, CoF2_Correa} and which has been extensively studied in the past due to its strong piezomagnetic effect~\cite{Borovik-Romanov_1960, Kharchenko_1985, Kharchenko_1995}. Being isostructural to RuO$_2$, CoF$_2$ is an altermagnet with an experimentally measured local magnetization of $2.21\,\mathrm{\mu_B}$~\cite{Strempfer_2004}, although the estimated values of the band-gap and of the magnetocrystalline anisotropy energy strongly depend on the first-principles Hubbard parameter $U$~\cite{CoF2_Correa}. Due to their structural inversion symmetry, RuO$_2$ and CoF$_2$ allow for linear in an electric field currents like anomalous and spin/orbital Hall effects but do not allow for second order photocurrents. 

On the other hand, owing to different symmetry requirements, centrosymmetric crystals allow for spin and orbital induced moments, commonly known as the inverse Faraday effect~\cite{Battiato_2014, Berritta_2016}, and  optical torques on the spins~\cite{freimuth_2016}, which are  second order in applied electric field~\cite{Xiao_2023}. The photo-induced magnetism is an important mechanism for optical detection of magnetic order~\cite{Kirilyuk_2010}, while the optical spin torques have been shown to drive THz emission~\cite{Huisman_2016}, antiferromagnetic dynamics and switching~\cite{Ross_2023}. In case of antiferromagnets (AFMs), the inverse Faraday effect and optical torques have been studied recently for Mn$_2$Au \cite{Merte_Mn2Au, Freimuth_2021_Mn2Au}, however, very little is known about the photo-induced magnetism in altermagnetic materials.
Here, based on first principles methodology in combination with the Keldysh formalism for photoresponse, we study in detail the light-induced spin and orbital magnetism in rutile RuO$_2$ and CoF$_2$, and discuss the consequences of our findings for the optical control and detection of  altermagnetism.

{\bf Method}

We employ the Keldysh formalism~\cite{freimuth_2016, freimuth_2021,*freimuth2017laserinducedarxiv} in order to compute the second order induced orbital and spin magnetization with Cartesian component $\delta\mathcal{O}_i$, emerging as a response to a continuous laser pulse of frequency $\omega$ according to:
\begin{align}\label{eq:keldysh}
    \delta\mathcal{O}_{i}=-\frac{\hbar a_{0}^{3} I}{2 c}\frac{\mathcal{E}_{\mathrm{H}}}{\left( \hbar \omega\right)^{2}} \operatorname{Im} \sum_{j k} \epsilon_{j} \epsilon_{k}^{*} \varphi_{i j k},
\end{align}
where $a_0 = 4\pi \epsilon_0 \hbar^2 /(m_e e^2)$ is the Bohr's radius, $I = \epsilon_0 c E_0^2/2$ is the intensity of the pulse, $\epsilon_0$ is the vacuum permittivity, $m_e$ is the electron mass, $e$ is the elementary charge,  $\hbar$ is the reduced Planck constant, $c$ is the light velocity, $\mathcal{E}_H=e^2/(4\pi \epsilon_0 a_0)$ is the Hartree energy, and $\epsilon_j$ is the $j$-th component of the polarization vector of the pulse. The tensor $\varphi_{ijk}$ (see Eq.(14) of Ref.~\cite{freimuth_2016} for a detailed form) is expressed in terms of the Green functions of the system, two velocity operators ($v_i$ and $v_j$) and the $i$-th component of the orbital angular momentum (OAM) operator $L_{i}$, or the $i$-th component of the Pauli matrix $\sigma_i$. For the orbital response, the prefactor in Eq.~\eqref{eq:keldysh} must be multiplied by an additional factor of 2. Within the Keldysh formalism, a constant lifetime broadening $\Gamma$ is introduced to account for effects of disorder on the electronic states.

{\bf Symmetry analysis}

In order to determine the symmetry allowed components of the tensor $\varphi_{ijk}$ in Eq.~(\ref{eq:keldysh}) in a compensated AFM, we first need to find out if it transforms like an axial (a), polar (p), staggered axial (sa), or staggered polar (sp) tensor~\cite{Freimuth_2021_Mn2Au}. Then the symmetry allowed components can be found by applying the Neumann’s principle.

In the case of the altermagnets considered here we calculate the atom-resolved orbital/spin polarization and then we study separately the overall induced orbital/spin moment in the unit cell, defined as the sum of all induced moments in the unit cell, as well as the staggered component, defined as the difference between the moments on different sublattices: $\delta\mathcal{O}^{\text{stag}} = \frac{1}{2} \big[\delta\mathcal{O}(\text{Atom}_{1})-\delta\mathcal{O}(\text{Atom}_{2})\big]$. At first, we assume that the altermagnets studied here are collinear and fully compensated, which allows to expand the laser-induced response in orders of the N\'eel vector $\mathbf{N}$ as:
\begin{align}
    \delta\mathcal{O}_{i} = & \ \varphi_{ijk}^{(3a)} E_{j}E^{\text{*}}_{k} + \varphi_{ijkl}^{(4sp)} E_{j}E^{\text{*}}_{k}N_{l} \notag \\
    & + \varphi_{ijklm}^{(5a)} E_{j}E^{\text{*}}_{k}N_{l}N_{m} + \cdots ,
\end{align}
for the total responses and
\begin{align}
    \delta\mathcal{O}_{i}^{stag} = & \ \varphi_{ijk}^{(3sa)} E_{j}E^{\text{*}}_{k} + \varphi_{ijkl}^{(4p)} E_{j}E^{\text{*}}_{k}N_{l} \notag \\
    & + \varphi_{ijklm}^{(5sa)} E_{j}E^{\text{*}}_{k}N_{l}N_{m} + \cdots ,
\end{align}
for the staggered responses. Above, $\varphi_{ijk}^{(3a)/(3sa)}$ is a non-staggered/staggered axial tensor of rank 3 that is independent of the staggered magnetization, $\varphi_{ijk}^{(4p)/(4sp)}$ is a non-staggered/staggered polar tensor of rank 4 that is odd in the staggered magnetization, and $\varphi_{ijk}^{(5a)/(5sa)}$ is a non-staggered/staggered axial tensor of rank 5 that is even in the staggered magnetization.  

For the $P4_2/mnm$ rutile structure of RuO$_2$ and CoF$_2$ we find out that 3 nonstaggered and 3 staggered axial tensors of rank 3 are allowed, which are listed in Table~\ref{table_1}, together with 10 staggered and 11 nonstaggered polar tensors of rank 4, which are listed in Table~\ref{table_2}. Additionally, our symmetry analysis reveals the existence of 30 nonstaggered and 30 staggered axial tensors of rank 5 that are allowed but not listed here for simplicity, even though we refer to them in the main text. In Tables~\ref{table_1} and~\ref{table_2} we use the notation below to refer to the respective tensors of the 4th rank:
\begin{align}
    \delta_{nopq}^{(ijkl)} = \delta_{in}\delta_{jo}\delta_{kp}\delta_{lq} \rightarrow \langle ijkl \rangle,
\end{align}
and analogously for the tensors of 3rd and 5th rank. In the following, we compare the results of our ab-initio calculations  to the predictions of the symmetry analysis.

\begin{table}[t!]
    \centering
    \begin{tabular}{c c c c c c}
    \hline\hline
   & & & & & \\
     \# & $\varphi_{ijk}^{(3a)}$ & Remark & \# & $\varphi_{ijk}^{(3sa)}$ & Remark \\ [0.5ex]
     \hline \\
     1 & $\langle$xzy$\rangle$ - $\langle$yzx$\rangle$ & $\checkmark$ & 4 & $\langle$yzy$\rangle$ - $\langle$xzx$\rangle$ & $\checkmark$ \\ [0.5ex]
     2 & $\langle$zxy$\rangle$ - $\langle$zyx$\rangle$ &       & 5 & $\langle$yyz$\rangle$ - $\langle$xxz$\rangle$ & \{4\} \\ [0.5ex]
     3 & $\langle$xyz$\rangle$ - $\langle$yxz$\rangle$ & \{1\} & 6 & $\langle$zyy$\rangle$ - $\langle$zxx$\rangle$ &       \\ [0.5ex]
    \hline\hline
    \end{tabular}
    \caption{Nonstaggered $\varphi_{ijk}^{(3a)}$ and staggered $\varphi_{ijk}^{(3sa)}$ axial tensors of rank 3. $\checkmark$: marks the tensors which correctly predict the presented components. A number in \{\}-brackets specifies the tensor that can be used as a replacement.}
    \label{table_1}
\end{table}

\begin{table}[t!]
    \centering
    \begin{tabular}{c c c c c c}
    \hline\hline
   & & & & & \\
     \# & $\varphi_{ijkl}^{(4sp)}$ & Remark & \# & $\varphi_{ijkl}^{(4p)}$ & Remark \\ [0.5ex]
     \hline \\
     7 & $\langle$yyyx$\rangle$ + $\langle$xxxy$\rangle$ & $\parallel$ & 17& $\langle$zzzz$\rangle$                  & $\checkmark\perp$ \\ [0.5ex]
     8 & $\langle$xxyx$\rangle$ + $\langle$yyxy$\rangle$ & $\checkmark\parallel$ & 18& $\langle$xyyx$\rangle$ + $\langle$yxxy$\rangle$ & $\parallel$ \\ [0.5ex]
     9 & $\langle$xyxx$\rangle$ + $\langle$yxyy$\rangle$ & \{8\} & 19& $\langle$yxyx$\rangle$ + $\langle$xyxy$\rangle$ & $\checkmark\parallel$ \\ [0.5ex]
     10& $\langle$yxxx$\rangle$ + $\langle$xyyy$\rangle$ & $\parallel$ & 20& $\langle$xxxx$\rangle$ + $\langle$yyyy$\rangle$ & $\parallel$ \\ [0.5ex]
     11& $\langle$yzzx$\rangle$ + $\langle$xzzy$\rangle$ & $\parallel$ & 21& $\langle$xzzx$\rangle$ + $\langle$yzzy$\rangle$ & $\parallel$ \\ [0.5ex]
     12& $\langle$yzxz$\rangle$ + $\langle$xzyz$\rangle$ & $\perp$ & 22& $\langle$xzxz$\rangle$ + $\langle$yzyz$\rangle$ & $\perp$ \\ [0.5ex]
     13& $\langle$zzyx$\rangle$ + $\langle$zzxy$\rangle$ & $\checkmark\parallel$ & 23& $\langle$xxzz$\rangle$ + $\langle$yyzz$\rangle$ & \{22\} \\ [0.5ex]
     14& $\langle$zyzx$\rangle$ + $\langle$zxzy$\rangle$ & \{14\} & 24& $\langle$yyxx$\rangle$ + $\langle$xxyy$\rangle$ & \{19\} \\ [0.5ex]
     15& $\langle$zyxz$\rangle$ + $\langle$zxyz$\rangle$ & $\checkmark\perp$ & 25& $\langle$zzxx$\rangle$ + $\langle$zzyy$\rangle$ & $\checkmark\parallel$ \\ [0.5ex]
     16& $\langle$yxzz$\rangle$ + $\langle$xyzz$\rangle$ & \{12\} & 26& $\langle$zxzx$\rangle$ + $\langle$zyzy$\rangle$ & \{25\} \\ [0.5ex]
       &                                 &  & 27& $\langle$zxxz$\rangle$ + $\langle$zyyz$\rangle$ & $\checkmark\perp$ \\ [0.5ex]
    \hline\hline
    \end{tabular}
    \caption{Staggered $\varphi_{ijkl}^{(4sp)}$ and nonstaggered $\varphi_{ijkl}^{(4p)}$ polar tensors of rank 4. $\checkmark$: marks the tensors which correctly predict the presented components. $\parallel$: The response is nonzero when $\mathbf{N}$ lies in the $xy$ plane. $\perp$: The response is nonzero when $\mathbf{N}$ points in the $z$ direction. A number in \{\}-brackets specifies the tensor that can be used as a replacement.}
    \label{table_2}
\end{table}

{\bf Results} 

{\bf Non-relativistic $\mathbf{RuO_2}$} 

\begin{figure*}[t!]
\begin{center}
\rotatebox{0}{\includegraphics [width=0.98\linewidth]{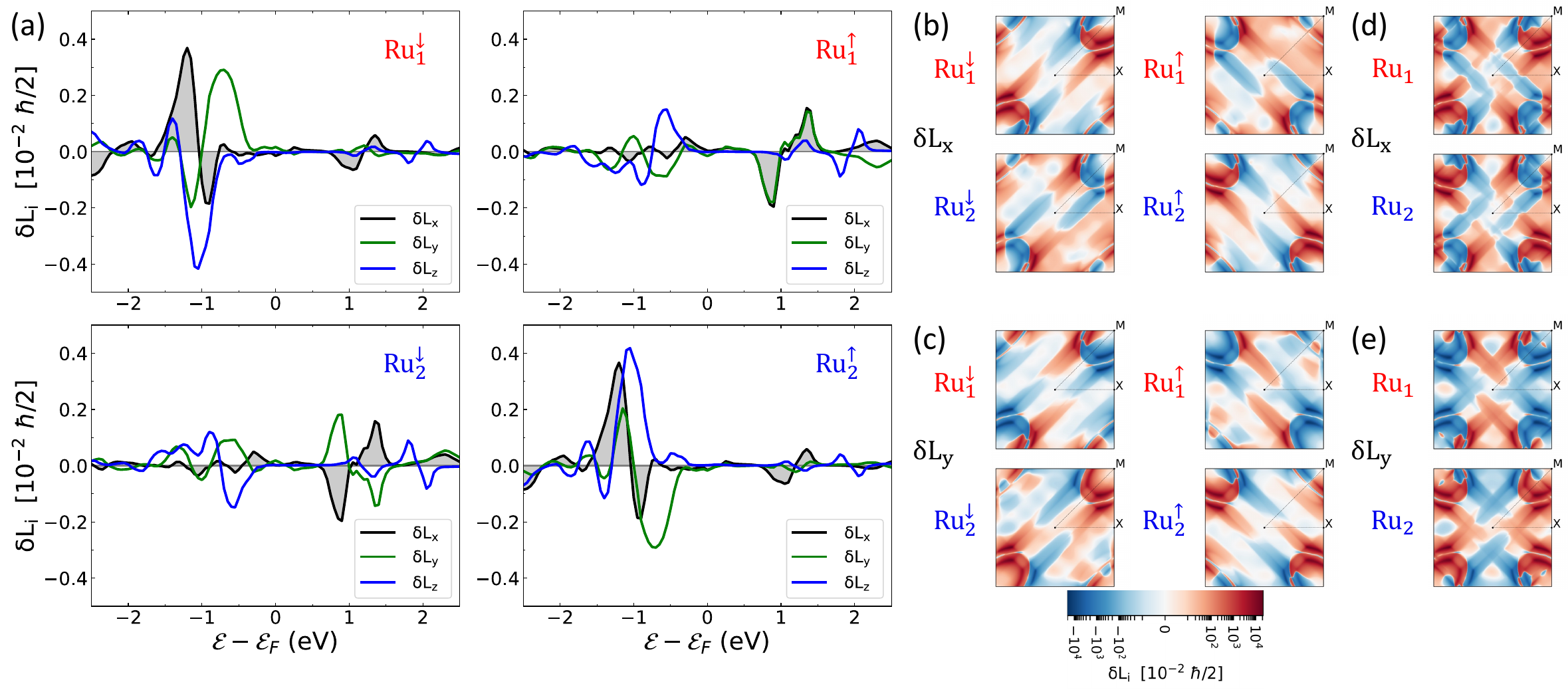}}
\end{center}
\caption{{\bf Non-collinear spin- and atom-resolved laser-induced orbital magnetization in non-relativistic $\mathbf{RuO_2}$}. (a) Cartesian components of the laser-induced orbital magnetization in relation to the band filling for  non-relativistic RuO$_2$. The spin-resolved ($\uparrow$ vs $\downarrow$) response is further decomposed into the contributions from two Ru atoms (Ru$_1$ vs Ru$_2$).  Among the responses, only $\delta L_x$ (shaded) survives in a system without oxygen atoms, which corresponds to a $\mathcal{TS}$ AFM simple tetragonal lattice of Ru atoms.  
(b-e): Reciprocal space distribution of laser-induced orbital magnetization in non-relativistic RuO$_2$.
(b-c): Non-staggered $\delta L_x$ (b) and staggered $\delta L_y$ (c) components decomposed into Ru contributions for each spin channel. (d-e): Non-staggered $\delta L_x$ (d) and staggered $\delta L_y$ (e) components decomposed into Ru contributions and summed over spin directions. The reciprocal space distributions are plotted over the constant energy surface at the true Fermi level $\mathcal{E}_{F}$.  In all calculations, the light frequency is $\hbar\omega=0.25$\,eV, lifetime broadening  $\Gamma=25$\,meV, light is polarized along [011] direction and its intensity is $I=10$\,GW/cm$^2$.}
\label{Fig2}
\end{figure*}

First, we explore purely altermagnetic effects, which are not influenced by the presence of spin-orbit coupling (SOC). We begin our discussion by examining the case of non-relativistic RuO$_2$. The non-relativistic bandstructure of this compound, presented in Fig.~\ref{Fig1}(a), clearly reflects the large altermagnetic splittings which occur between the spin-up and spin-down states along the M--$\Gamma$ and A--Z high-symmetry paths. The crystal structure with the magnetic moments pointing along the in-plane [110] direction is shown in Fig.~\ref{Fig1}(g). As without SOC only orbital photo-response is allowed, in Fig.~\ref{Fig2}(a) we present the results for the orbital magnetization which is induced by light linearly polarized along the [011] direction in relation to the band filling, decomposed into Ru  and spin contributions. At first glance we observe that all corresponding magnetization components are roughly of the same magnitude, leading to an induction of a strongly non-collinear, and not simply staggered  response  at each Ru atom. The pronounced peaks which appear in the signal can be easily traced back to the altermagnetic splittings of the bands in the energy regions of $[-1.5, -0.5]$\,eV and $[+1.0, +2.0]$\,eV. Additionally, for the $[-1.5, -0.5]$\,eV region there is a contribution originating from flat bands visible along the $\Gamma$--Z--R--A--Z path.


The emergence of these specific non-relativistic components can be very well understood from the symmetry analysis for the tensor elements even in $\mathbf{N}$. This analysis, summarized in Table~\ref{table_1} and~\ref{table_2} above, predicts that for light linearly polarized along the [011] direction, $\delta L_x$ is nonstaggered as reflected in the  3rd rank tensor \#1, whereas $\delta L_y$ is staggered in accord to the 3rd rank tensor \#4. The $\delta L_z$ component is not captured by neither 3rd nor 4th rank tensors, although it is listed in both nonstaggered axial, e.g., $\langle$zzyxz$\rangle$ - $\langle$zzxyz$\rangle$, and staggered axial, e.g., $\langle$zyzyz$\rangle$ - $\langle$zxzxz$\rangle$, 5th rank tensors. Our atom-resolved analysis shows that $\delta L_z$ is staggered, therefore it is predicted by the 5th rank staggered tensor. We also note that the $\delta L_z$ response remains unchanged when the crystal is irradiated with light circularly-polarized in the [011] plane (not shown).

The coexistence of staggered and non-staggered components clearly presents a manifestation of the altermagnetic behavior which is expected to incorporate ferromagnetic and antiferromagnetic features. To demonstrate this point more clearly, we perform additional calculations for the orbital non-relativistic photoresponse of the same crystal with oxygen atoms removed. We observe that for this artificial $\mathcal{TS}$ symmetric $-$ where $\mathcal{S}$ is spatial inversion $\mathcal{P}$ or translation $\mathcal{\tau}$ and  $\mathcal{T}$ is time reversal $-$ simple tetragonal AFM lattice of Ru atoms the only surviving component is the non-staggered $\delta L_x$ [shaded black line in Fig.~\ref{Fig2}(a)], while the staggered components vanish. Additionally, we notice that the  components of $\delta\mathbf{L}$ are connected by the following relations
\begin{align}\label{eq:components_relation}
    \delta L^{\mathrm{Ru}^{\downarrow (\uparrow)}_{1}}_{x} = \delta L^{\mathrm{Ru}^{\uparrow (\downarrow)}_{2}}_{x} \quad \text{and} \quad
    \delta L^{\mathrm{Ru}^{\downarrow (\uparrow)}_{1}}_{y,z} = -\delta L^{\mathrm{Ru}^{\uparrow (\downarrow)}_{2}}_{y,z},
\end{align}
for light linearly polarized along [011]. 

The first relation in Eq.~\eqref{eq:components_relation} can be proven as follows:
\begin{equation}\label{eq_balance_nonstag}
    \delta L^{\mathrm{Ru}^{\uparrow}_{1}}_{x} +
    \delta L^{\mathrm{Ru}^{\downarrow}_{1}}_{x} =
    \delta L^{\mathrm{Ru}^{\uparrow}_{2}}_{x} +
    \delta L^{\mathrm{Ru}^{\downarrow}_{2}}_{x} 
\end{equation}
is a relation that follows from the fact that for light polarization along [011] the response
$L_x$ is non-staggered.
A second relation follows from the observation that the total response from all the wavefunctions with
spin $\uparrow$ has to be equal to the total response from all the wavefunctions with spin $\downarrow$
in this compensated altermagnet computed without SOC:
\begin{equation}\label{eq_balance_compensated}
    \delta L^{\mathrm{Ru}^{\uparrow}_{1}}_{x}+ 
    \delta L^{\mathrm{Ru}^{\uparrow}_{2}}_{x} 
    =
    \delta L^{\mathrm{Ru}^{\downarrow}_{1}}_{x} +
    \delta L^{\mathrm{Ru}^{\downarrow}_{2}}_{x}. 
\end{equation}
Subtracting Eq.~\eqref{eq_balance_compensated} from Eq.~\eqref{eq_balance_nonstag} and simplifying
the result we obtain the first relation in Eq.~\eqref{eq:components_relation}.

In order to obtain the second relation in Eq.~\eqref{eq:components_relation} we consider
the relation
\begin{equation}
    \delta L^{\mathrm{Ru}^{\uparrow}_{1}}_{y}+ 
    \delta L^{\mathrm{Ru}^{\downarrow}_{1}}_{y}- 
    \delta L^{\mathrm{Ru}^{\uparrow}_{2}}_{y} -
    \delta L^{\mathrm{Ru}^{\downarrow}_{2}}_{y}=0, 
\end{equation}
which states that the staggered response is zero in this simple crystal structure where the
oxygens have been removed. A second relation relevant in this case
is
\begin{equation}
    \delta L^{\mathrm{Ru}^{\uparrow}_{1}}_{y}+ 
    \delta L^{\mathrm{Ru}^{\downarrow}_{1}}_{y}+
    \delta L^{\mathrm{Ru}^{\uparrow}_{2}}_{y} +
    \delta L^{\mathrm{Ru}^{\downarrow}_{2}}_{y}=0. 
\end{equation}
This relation states that the total axial response along $y$ is zero for the light polarization
here considered.
Subtracting these latter two relations from each other we obtain the second part of Eq.~\eqref{eq:components_relation} for the axial response along $y$. The axial response along $z$ can be
discussed analogously.

\begin{figure*}[t!]
\begin{center}
\rotatebox{0}{\includegraphics [width=0.65\linewidth]{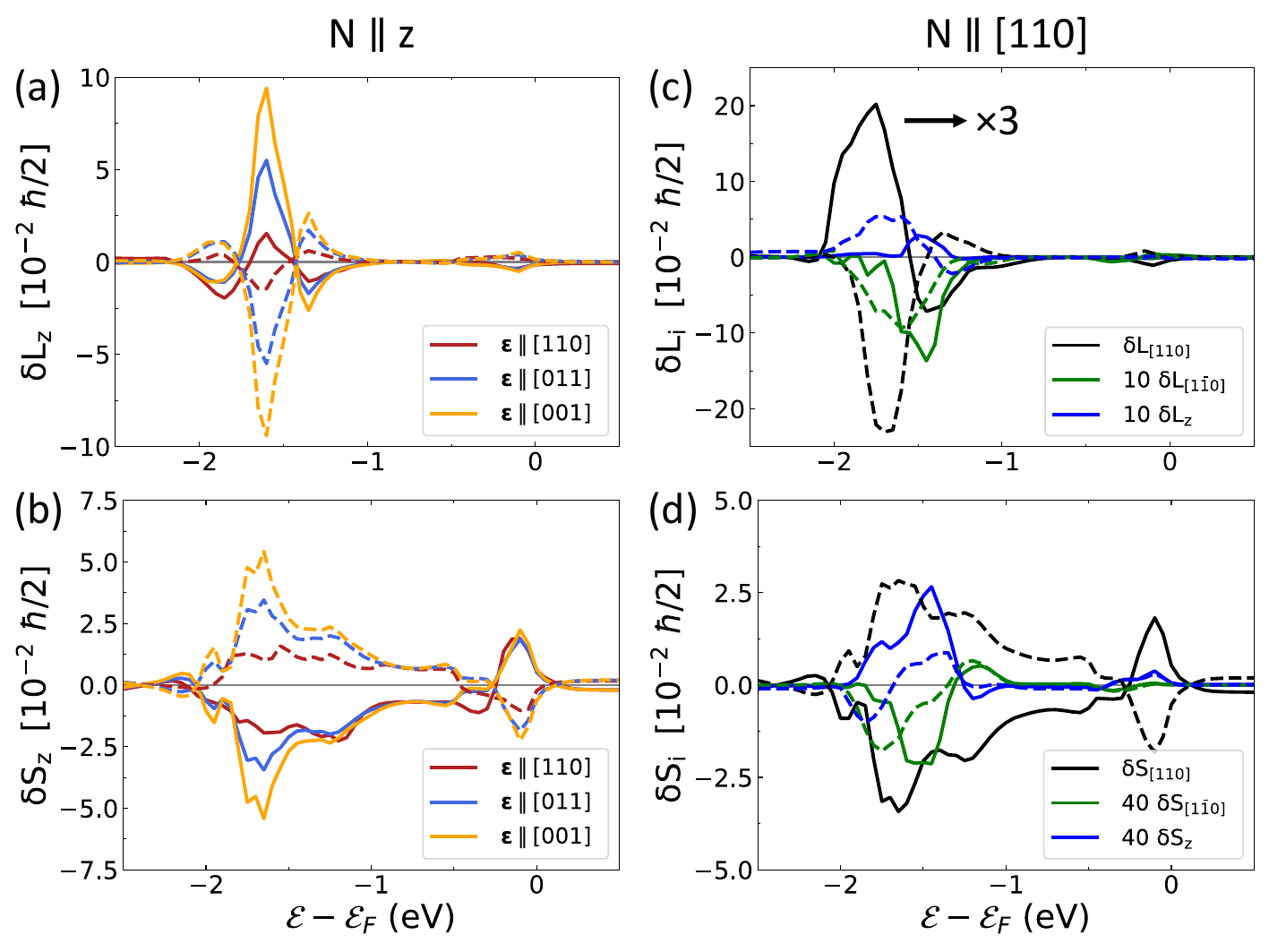}}
\end{center}
\caption{{\bf Atom-projected laser-induced orbital and spin magnetization in relativistic  $\mathbf{CoF_2}$}. (a-b): Orbital (a) and spin (b) $z$-components of the laser-induced magnetization as a function of the band filling, arising for different cases of linearly polarized light, in CoF$_2$ with $\mathbf{N}$ along the $z$ axis. In (a-b) the $x$ and $y$ magnetization components are not sizeable and are not shown. (c-d) Orbital (c) and spin (d) laser-induced magnetization in relation to the band filling, arising for linearly polarized light along the [011] direction, in the case of CoF$_2$ with $\mathbf{N}$ along the [110] direction. Note that in (c-d) the components which are perpendicular to $\mathbf{N}$ are significantly enlarged for comparison. In (a-d) the responses are depicted with either a solid or dashed line for Co$_1$ and Co$_2$ atoms, respectively. In all calculations, a light energy of $\hbar\omega=0.25$\,eV, a lifetime broadening  $\Gamma=25$\,meV and a light intensity of $I=10$\,GW/cm$^2$ were used.}
\label{Fig4}
\end{figure*}

The above relations hold true even after summing the contributions of each spin direction on the respective Ru atom (not shown), which is expected from the non-staggered origin of $\delta L_x$ and the staggered origin of $\delta L_y$ and $\delta L_z$. Similar relations also hold for linear polarization along [101] (not shown), with  $\delta L_y$ assuming the role of the non-staggered component, while the staggered components are $\delta L_x$ and $\delta L_z$, following the symmetry of  tensors \#1 and \#4. 
Additionally, both our symmetry analysis and our calculations reveal further relations which connect the non-vanishing components for different directions of the linear light polarization vectors:
\begin{align}
    \delta L_{x}^{\mathrm{Ru}_{1}^{\downarrow(\uparrow)}/\mathrm{Ru}_{2}^{\downarrow(\uparrow)}}, \epsilon \parallel [011] &= - \delta L_{y}^{\mathrm{Ru}_{1}^{\downarrow(\uparrow)}/\mathrm{Ru}_{2}^{\downarrow(\uparrow)}}, \epsilon \parallel [101] 
    \label{equation1}\\
    \delta L_{y}^{\mathrm{Ru}_{1}^{\downarrow(\uparrow)}/\mathrm{Ru}_{2}^{\downarrow(\uparrow)}}, \epsilon \parallel [011] &= - \delta L_{x}^{\mathrm{Ru}_{1}^{\downarrow(\uparrow)}/\mathrm{Ru}_{2}^{\downarrow(\uparrow)}}, \epsilon \parallel [101] 
    \label{equation2}\\
    \delta L_{z}^{\mathrm{Ru}_{1}^{\downarrow(\uparrow)}/\mathrm{Ru}_{2}^{\downarrow(\uparrow)}}, \epsilon \parallel [011] &= - \delta L_{z}^{\mathrm{Ru}_{1}^{\downarrow(\uparrow)}/\mathrm{Ru}_{2}^{\downarrow(\uparrow)}}, \epsilon \parallel [101].
    \label{equation3}
\end{align}
Overall, we predict that in the non-relativistic case the optically induced orbital magnetization in rutile altermagnets will have a strongly canted character on neighboring magnetic ions, with the direction and sense of the relative canting controllable by the light polarization. We emphasize that this is a purely altermagnetic effect, as it is absent in $\mathcal{TS}$ AFMs.

To understand better the peculiar symmetry of the orbital response, next, we examine its behavior in the reciprocal space. In Fig.~\ref{Fig2}(b-c) we present the Ru- and spin-resolved distribution of the induced components $\delta L_x$ and $\delta L_y$ in Fig.~\ref{Fig2}(b) and Fig.~\ref{Fig2}(c), respectively, as well as their sum over spin  in Fig.~\ref{Fig2}(d) and Fig.~\ref{Fig2}(e). The distributions are plotted at the true Fermi level for $k_z=0$, and roughly consist of equal contributions with opposite sign $-$ a behavior which explains their suppression at this energy when integrated over $k$-space, see Fig~\ref{Fig2}(a). First of all, all distributions obey an inversion symmetry with respect to $\Gamma$ and a $C_{2z}$ symmetry with respect to a $\pi$-rotation around $z$. Moreover, we notice that if we apply a $[C_{4z}||M_{[110]}]$ or $[C_{4z}||M_{[1\bar{1}0]}]$ symmetry operation, where $C_{4z}$ is a 4-fold rotation around the $z$-axis and $M_{[110]}$ ($M_{[1\bar{1}0]}$) is a mirror symmetry with respect to a plane perpendicular to the $[110]$ $([1\bar{1}0])$ diagonal, the $\delta L_x$ distribution of the Ru$_{1}^{\downarrow}$ atom becomes equal to the distribution of the Ru$_{2}^{\uparrow}$ atom, with the same also applying for the Ru$_{1}^{\uparrow}$ and Ru$_{2}^{\downarrow}$ atoms. For the $\delta L_y$ distributions, apart from the aforementioned symmetry operations, a change of sign is also needed to relate the respective atoms. This means  that the relations of Eq.~\eqref{eq:components_relation} hold true in the reciprocal space. We note that the $C_{4z}$ rotation is responsible for interchanging the Ru atoms between the two opposite spin sublattices in RuO$_2$. Additionally, after summation over the spins in Figs.~\ref{Fig2}(d-e), the aforementioned symmetry operations connect the response on the two Ru atoms, in the same way as Eq.~\eqref{eq:components_relation} holds true after spin summation.


{\bf $\mathbf{CoF_2}$}

We continue our discussion by considering the case of altermagnetic rutile CoF$_2$. In Fig.~\ref{Fig1}(c) we present a non-relativistic bandstructure and in Fig.~\ref{Fig1}(d) the bandstructure computed with SOC for the experimentally observed configuration with $\mathbf{N} \Vert z$. We observe a very small difference between the relativistic and the non-relativistic bands on the quantitative level of band dispersions. However, we point out that the band representations induced by the relativistic magnetic and nonrelativistic spin symmetries \cite{Smejkal_PRX_2022a} are in general different.  For instance, the the band representation at R is double (single) dimensional for the nonrelativistic (relativistic) case. The calculated electronic structure with a bandgap of $\sim2.0$\,eV and magnetic moments of 2.66$\mathrm{\mu_B}$ is in accordance with previous first-principles calculations~\cite{CoF2_Correa}. The altermagnetic splittings, which are  again located along the $\Gamma$--M and A--Z high-symmetry paths, are much smaller in magnitude  in the case of CoF$_2$ as compared to RuO$_2$.
First we calculate the laser-induced orbital and spin magnetization for the case of relativistic CoF$_2$ with $\mathbf{N}\Vert z$,  presenting the results in Fig.~\ref{Fig4}(a) and Fig.~\ref{Fig4}(b), respectively, in relation to the band-filling. The responses are projected onto each Co atom and are drawn with either a solid or dashed line for each one of them. We consider different cases of light polarization and discuss only the $z$-component as the $x,y$-components are more than one order of magnitude smaller. 

We observe that both orbital and spin responses become large predominantly in the energy region of $[-2.0, -1.0]$\,eV, with the orbital signal being generally almost two times larger and much sharper in energy. This is rather expected given that in this energy region a large number of Co $d$-states is located giving rise to altermagnetic splittings and flat bands. 
The highest peak occurs at $\sim-1.6$\,eV reaching values as high as $10^{-1}\hbar/2$ for the induced orbital and $5.0\cdot10^{-2}\hbar/2$ for the induced spin moments. 
We remark that these values may increase even further upon reduction in the  broadening value in ultra-clean systems~\cite{Mu_2021, Zhou_2022, Merte_Mn2Au}. 
A closer view at the energy region of $[-2.0, -1.0]$\,eV reveals that the sharp peaks in the responses occur because the used laser frequency of $\hbar\omega=0.25$\,eV is similar to the size of the altermagnetic splittings of CoF$_2$. The fact that the orbital response peaks in this energy range underlines the crystal-field origin of the altermagnetic spin splitting. Correspondingly, a shift in the light frequency to higher values results in a suppression of the orbital signal.  We also notice the existence of a smaller peak at $\sim-0.2$\,eV which is larger for the spin response and can be attributed to the small splitting appearing around this energy.

The case of $\mathbf{N} \Vert z$ is highly symmetric with perfectly compensated staggered magnetization, which is also known not to support the anomalous Hall effect~\cite{Smejkal_2020}. In this case, our symmetry analysis is in perfect accord to the ab-initio calculations. The presented component in response to linear polarization along the [001] direction [yellow line in Fig.~\ref{Fig4}(a-b)] is staggered, odd in the magnetization, as predicted by the 4th rank tensor \#17. The component which arises for light linearly polarized along the [110] direction [red line in Fig.~\ref{Fig4}(a-b)] appears both as nonstaggered and staggered. The nonstaggered part is odd in the magnetization and predicted by the 4th rank tensor \#15. We note that for circular polarization in the [110] plane (not shown), the nonstaggered part is even in the magnetization and predicted by the 3rd rank tensor \#2. On the contrary, the staggered part is odd in the magnetization and predicted by the 4th rank tensor \#27. Furthermore, when light is linearly polarized along the [011] direction [blue line in Fig.~\ref{Fig4}(a-b)], a staggered component arises which is odd in the magnetization and occurs as a linear combination of the 4th rank tensors \#17 and \#27.


Symmetry-wise, the case of CoF$_2$ with $\mathbf{N}\Vert$[110] presents a more complex behavior, Fig.~\ref{Fig4}(c-d),  as it allows for ferromagnetic magnetization, as discussed below. When we compare the orbital and spin induced moments in Fig.~\ref{Fig4}(c-d) to those arising in response to linearly polarized light along the [011] direction for $\mathbf{N}\Vert z$, we observe that among all Cartesian components the orbital signal is two times larger and the spin signal is somewhat smaller, with peaks appearing in the same energy regions. Moreover, we find that the two components perpendicular to $\mathbf{N}$ are small but not vanishing, and we plot them after multiplying with a large factor to enable comparison of the qualitative behavior of all components. Our symmetry analysis, where we expand
the tensors only in orders of $\mathbf{N}$, predicts the components parallel and perpendicular to $\mathbf{N}$ [black and green lines in Fig.~\ref{Fig4}(c-d), respectively] to be non-staggered and staggered at the same time, while being even in the magnetization and described by 5th rank tensors, e.g., nonstaggered axial $\langle$yyzxy$\rangle$ - $\langle$xxzyx$\rangle$ and staggered axial, $\langle$yzxyx$\rangle$ - $\langle$xzyxy$\rangle$. Our calculations confirm the simultaneous non-staggered and staggered nature of these components but reveal them to be odd in the magnetization, in contrast to the symmetry expectations. On the other hand, we find the out-of-plane component [blue line in Fig.~\ref{Fig4}(c-d)] to be non-staggered and staggered at the same time, odd in the magnetization, and described by the 4th rank tensors \#13 and \#25, in full agreement with our symmetry analysis.

\begin{figure*}[t!]
\begin{center}
\rotatebox{0}{\includegraphics [width=0.90\linewidth]{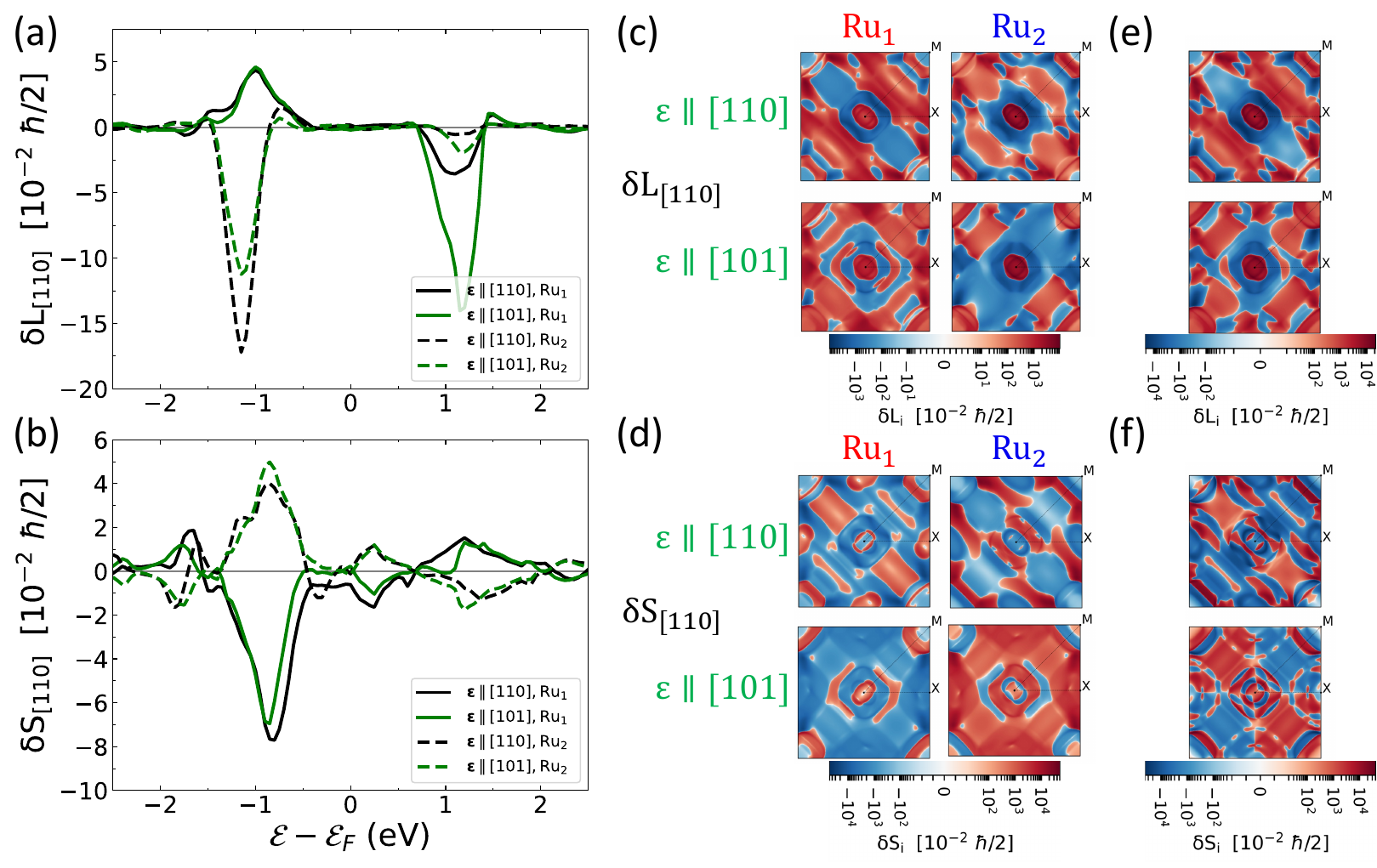}}
\end{center}
\caption{{\bf Spin and orbital photoresponse in relativistic  $\mathbf{RuO_2}$ with $\mathbf{N\Vert[110]}$}. (a-b): Orbital (a) and spin (b) laser-induced magnetization in relation to the band filling, arising for linearly polarized light along the [110] or the [101] directions, in the case of RuO$_2$ with SOC and $\mathbf{N}$ along [110]. (c-f): Reciprocal space distribution of laser-induced orbital and spin magnetization, at the shifted Fermi energy level $\mathcal{E}_{F}^{\prime}=\mathcal{E}_{F}-1.0$\,eV. (c-d): Orbital (c) and spin (d) Ru-atom-resolved moments. (e-f): Orbital (e) and spin (f) moments summed on all atoms. 
In (a-f) only the component parallel to $\mathbf{N}$ is sizeable and presented. In (a-b) the responses on the respective Ru atom are depicted with either a solid or dashed line. In (c-d) the left (right) plots present the responses projected on the Ru$_{1}$ (Ru$_2$) atom. In (c-f) the upper (lower) plots present the responses arising for linearly polarized light along the [110] ([101]) direction. In all calculations, a light energy of $\hbar\omega=0.25$\,eV, a lifetime broadening  $\Gamma=25$\,meV and a light intensity of $I=10$\,GW/cm$^2$ were used.}
\label{Fig5}
\end{figure*}

{\bf Relativistic $\mathbf{RuO_2}$}

\begin{figure*}[t!]
\begin{center}
\rotatebox{0}{\includegraphics [width=0.92\linewidth]{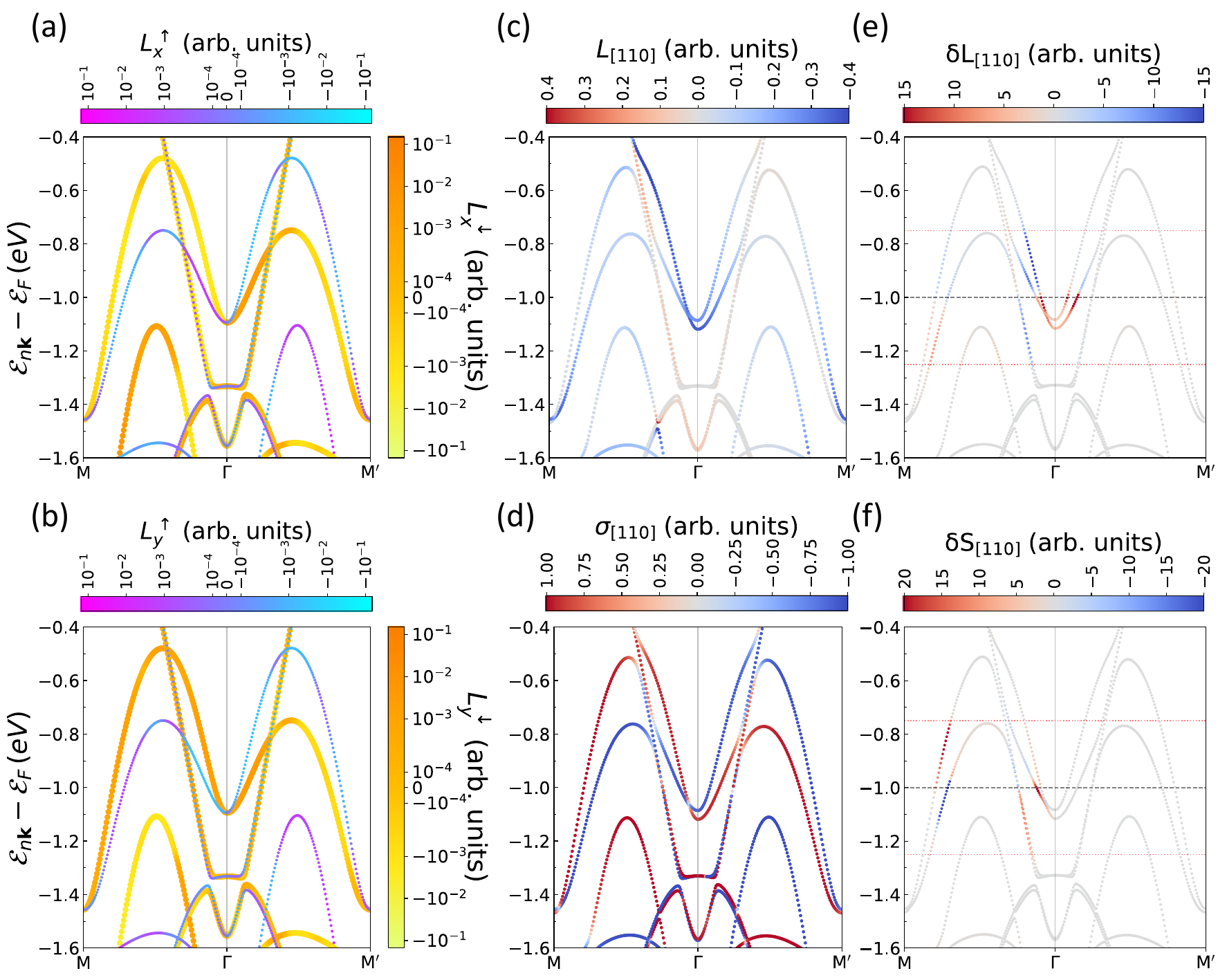}}
\end{center}
\caption{{\bf Band resolved laser-induced orbital and spin magnetization in $\mathbf{RuO_2}$}. (a-b): Altermagnetic spin-splittings along M--$\Gamma$--M$^\prime$ in the spin-resolved band-structure of non-relativistic RuO$_2$, colored by the value of (a) $L_x$ and (b) $L_y$ components of the OAM operator. (c-d): Corresponding band-structure of relativistic RuO$_2$ with $\mathbf{N}\Vert$[110] colored by the value of the (c) OAM and (d) spin projected onto [110]. (e-f): Band-resolved two-band contributions to the orbital (e) and spin (f) laser-induced moments, at the shifted Fermi energy level $\mathcal{E}_{F}^{\prime}=\mathcal{E}_{F}-1.0$\,eV. The presented components parallel to $\mathbf{N}$ arise for light linearly polarized along the [110] direction. The horizontal dotted red lines at $\mathcal{E}_{F}^{\prime}\pm0.25$\,eV denote the laser pulse energy. In (a-f): $\Gamma$(0, 0, 0), M(0.5, 0.5, 0) and M$^\prime$(0.5, -0.5, 0).
}
\label{Fig7}
\end{figure*}

We further explore the properties of spin and orbital photoresponse for $\mathbf{N}\Vert[110]$ by coming back to  RuO$_2$, where this situation is much closer to experimentally observed conditions~\cite{Berlijn_2017, Zhu_2019}. The main energy scales of the bandstructure for this case, in analogy to CoF$_2$, do not show remarkable deviations from the non-relativistic bands,  Fig.~\ref{Fig1}(b). The impact of SOC on the photoresponse in RuO$_2$ is, however, remarkable. In Fig.~\ref{Fig5}(a-b) we present Ru-resolved orbital and spin moments, which exhibit a very large component parallel to $\mathbf{N}$ for linear polarization along the [110] and [101] directions. After including  SOC into account, the responses reach values which are two orders of magnitude larger than in the non-relativistic case, Fig.~\ref{Fig2}(a), accounting to about $-1.75\cdot10^{-1}\hbar/2$ for the orbital and $-8.0\cdot10^{-2}\hbar/2$ for the spin channels, developing strong peaks around $\sim-1.0$\,eV and $\sim1.2$\,eV, which originate in altermagnetic spin splitting. These values stand as colossal, when compared to those due to inverse Faraday effect in much heavier ferromagnets, such as FePt~\cite{Berritta_2016,freimuth2017laserinduced}, where they reach similar magnitude. The overall range of values induced by SOC is comparable for CoF$_2$ and RuO$_2$, which underlines the fact that in contrast to ferromagnets, it is not the strength of the spin-orbit interaction which drives the prominent response.

The orbital response is very different for each Ru atom and also varies for each polarization case. On the other hand, the spin response varies slightly among ruthenia and remains rather insensitive to the change in light polarization. The case with light polarization along [110] [black line in Fig.~\ref{Fig5}(a-b)] is correctly predicted by our symmetry analysis with the response being non-staggered and staggered at the same time, as described by the 4th rank tensors \#8 and \#19, and odd in the staggered magnetization. On the contrary, the case with light polarization along the [101] direction [green line in Fig.~\ref{Fig5}(a-b)] disagrees with our symmetry predictions because even though it is non-staggered and staggered at the same time, as expected, it is odd in the staggered magnetization and not even. This behavior with respect to the staggered magnetization violates the prediction of the 5th rank tensors, e.g., nonstaggered axial $\langle$yyzxy$\rangle$ - $\langle$xxzyx$\rangle$ and staggered axial, $\langle$yzxyx$\rangle$ - $\langle$xzyxy$\rangle$ components. 

In order to get further insight into the microscopic behavior of the photo-induced moments, we analyze their anatomy in the reciprocal space. In Fig.~\ref{Fig5}(c-d) we present the Brillouin zone distribution of the orbital and spin moments, respectively, separately for each Ru atom, and for two different cases of linear polarization. We focus on the component which is parallel to $\mathbf{N}$ and calculate the distributions for the shifted Fermi energy level $\mathcal{E}_{F}^{\prime}=\mathcal{E}_{F}-1.0$\,eV around which the responses in Fig.~\ref{Fig5}(a-b)  peak. The induced moments in Fig.~\ref{Fig5}(c-d), similarly to the non-relativistic scenario in Fig.~\ref{Fig2}(d-e), generally consist of large uniform regions of either positive or negative contributions. Qualitatively, we observe that the spin moments in Fig.~\ref{Fig5}(d) exhibit a sign reversal, when compared to the orbital moments in Fig.~\ref{Fig5}(c), which is consistent with the integrated responses in Fig.~\ref{Fig5}(a-b). Moreover, a change of sign also takes place when comparing the distributions between the two Ru atoms [left (right) plots for Ru$_1$ (Ru$_2$) in Fig.~\ref{Fig5}(c-d)], which is especially visible for spin, but much less obvious for the case of the orbital response. This results in very different magnitude of the orbital peaks in the opposite sublattices.

Locally, the spin response exhibits hot spots where its magnitude is larger than the orbital response by an order of magnitude, however, this does not result in a larger overall spin response when integrated over the Brillouin zone. In fact, when the photoresponses are summed over all atoms in the unit cell in Fig.~\ref{Fig5}(e-f), the orbital distribution in Fig.~\ref{Fig5}(e) remains rather uniform compared to the spin counterpart in Fig.~\ref{Fig5}(f), which exhibits a finer structure  due to the spin-splittings brought by SOC having a much weaker effect on the local distribution of orbital moments. In all cases we notice a strong dependence on the light polarization, with drastically different distributions appearing for linear polarization along  [110] [upper plots in Fig.~\ref{Fig5}(c-f)] or [101] [lower plots in Fig.~\ref{Fig5}(c-f)]. In this relativistic scenario, each distribution in Fig.~\ref{Fig5}(c-f) still exhibits a $C_{2z}$ rotation and inversion with respect to $\Gamma$ symmetries in the reciprocal space, but, as opposed to the non-relativistic case, the $[C_{4z}||M_{[110]}]$ symmetry is not even closely preserved. 

Further information concerning the microscopic behavior of the studied effects can be derived from the analysis of band-resolved contributions. In Fig.~\ref{Fig7}(a-b) we present a detailed view of the non-relativistic bands of RuO$_2$, in the vicinity of the $\mathcal{E}_{F}-1.0$\,eV energy level, where large altermagnetic splittings appear. For a given (up or down) spin state, we notice a pronounced asymmetry of the bands when moving from M--$\Gamma$ to $\Gamma$--M$^\prime$. An analysis of the orbital character of the states reveals that $L_x$ [Fig.~\ref{Fig7}(a)] and $L_y$ [Fig.~\ref{Fig7}(b)] are equal along the $\Gamma$--M$^\prime$ path and opposite along the M--$\Gamma$ path. Then, in the relativistic scenario, where the states become SOI-entangled, the orbital character in Fig.~\ref{Fig7}(c) is uniform and preserves the sign around $\Gamma$ point, while the spin character in Fig.~\ref{Fig7}(d) alternates sign. 

Finally, in Fig.~\ref{Fig7}(e-f) we examine the contribution from the resonant two-band transitions~\cite{Adamantopoulos_2022} to the band-resolved induced orbital and spin magnetization, respectively, and witness a drastically opposite behavior. While for the orbital response the transitions are mainly located at the center of the Brillouin zone, for the spin response they move further away to the edges.
Additionally, the orbital transitions preserve their sign while moving from M--$\Gamma$ to $\Gamma$--M$^\prime$, whereas the spin transitions are suppressed in the $\Gamma$--M$^\prime$ path. As a result, in case of orbital response, we observe in Fig.~\ref{Fig5}(e) a uniform feature centered around $\Gamma$  which strongly contributes to the overall orbital signal. On the other hand, spin response in the vicinity of $\Gamma$ is driven by isolated hot-spot contributions originating in crossings among bands of opposite spin hybridizing via SOI. This results in a more ``staggered" spin response in the center of the Brillouin zone, Fig.~\ref{Fig5}(f), in analogy to the case of spin photocurrents in non-centrosymmetric systems~\cite{Adamantopoulos_2024}, which results in an overall suppression of light-induced spin moments.

{\bf Beyond altermagnetism}

Finally, we come back to the observed inconsistencies between our ab-initio calculations and the predictions of our symmetry analysis. When $\mathbf{N}$ is along $z$, RuO$_2$ and CoF$_2$ are perfectly compensated antiferromagnets. However, as soon as $\mathbf{N}$ is tilted away from $z$ an additional ferromagnetic component $\delta \mathbf{M}$ is
induced. $\delta \mathbf{M}$ and $\mathbf{N}$ are related by a staggered polar 2-nd rank tensor, i.e., $\delta M_i=\chi^{(2p)}_{ij}N_j$. In RuO$_2$ and CoF$_2$ $\chi^{(2p)}_{ij}$ is required by symmetry to have the form
\begin{equation}\label{eq_chisp_cof2}
\chi^{(2p)}_{ij}\rightarrow
\langle
xy
\rangle
+
\langle
yx
\rangle.
\end{equation}
Consequently, when $\mathbf{N}$ is along [110], $\delta \mathbf{M}$ is along [110], too. We suggest to
refer to this case as \textit{ferri-altermagnetism}.
However, when $\mathbf{N}$ is along [100], $\delta \mathbf{M}$ is along [010] implying a small canting.
We suggest to denote this case as \textit{canted altermagnetism}.

\begin{table}[t!]
    \centering
    \begin{tabular}{c c c }
    \hline\hline
   & & \\
     Atom & Spin moment & Orbital moment \\ [0.5ex]
     \hline \\
     Ru$_1$ & 1.11979 &-0.00245 \\ [0.5ex]
     Ru$_2$ &-1.12354 &-0.01864 \\ [0.5ex]
     O$_{1(,2)}$ &-0.00420 &-0.00571 \\ [0.5ex]
     O$_{3(,4)}$ & 0.00359 &-0.00126 \\ [0.5ex]
    \hline\hline
    \end{tabular}
    \caption{Atom-resolved  spin and orbital moments in RuO$_2$ along $\mathbf{N}$ for $\mathbf{N}\Vert$[110] (the moments are given in units of $\mu_{\rm B}$).}
    \label{table_3}
\end{table}

The size of $\delta \mathbf{M}$ can be estimated most easily from calculations when $\mathbf{N}$ is along
[110], because in this case the $\delta \mathbf{M}$ along [110] implies that there is an additional ferromagnetic
component parallel to the spin quantization axis -- which can be easily extracted from the site-resolved orbital and spin magnetic moments.
In Table~\ref{table_3} we list these site-resolved spin and orbital magnetic moments. We observe that while the spin moments on ruthenia differ by less then 0.004\,$\mathrm{\mu_B}$, an order of magnitude larger difference comes from the orbital channel. Manifestly, the orbital moments on all atoms are aligned, which also explains the consistent sign around $\Gamma$ in Fig.~\ref{Fig7}(c). In this context, to underline the orbital origin of the ferromagnetic magnetization, one can refer to this case as {\it orbital} ferri-altermagnet. 

Noteworthily, Eq.~\eqref{eq_chisp_cof2} does not only apply to the ground-state magnetization, but also
to the laser-induced magnetizations. For example 
$\chi^{(2p)}_{ij}$ implies that the tensor \#1, i.e., $\langle xzy\rangle -\langle yzx\rangle$, which describes a non-staggered response, entails a staggered response
of the form $\langle yzy\rangle -\langle xzx\rangle$, which is tensor \#4.
Similar examples are the following pairs of tensors: \#12 and \#22; \#7 and \#18; \#8 and \#19; \#10 and \#20; \#11 and \#21; \#12 and \#22.

Note that $\chi^{(2p)}_{ij}$ predicts the ferri-altermagnetism and the canted altermagnetism 
only in the relativistic case. In the
non-relativistic calculation there are additional symmetries in the spin channel which
remove the ferri-altermagnetism and the canting leading to a perfectly compensated 
altermagnet. However, even in the non-relativistic case $\chi^{(2p)}_{ij}$ predicts
the laser-induced staggered orbital response to be related to the non-staggered one
in terms of symmetry. Therefore, $\chi^{(2p)}_{ij}$ predicts the coexistence of
staggered and non-staggered orbital responses even in the non-relativistic case -- in harmony with the discussion in the section ``Non-relativistic RuO$_2$".

The question arises of whether Eq.~\eqref{eq_chisp_cof2} is sufficient to explain all 
aspects of the magnetic structure when $\mathbf{N}$ is tilted away from the $z$ axis. Clearly, there
may be additional higher-order corrections. Consider, for example, the 3rd rank axial 
tensor \#2, which is given by $\langle zxy \rangle-\langle zyx\rangle$. Obviously, it
allows the coupling
\begin{equation}
\delta N_z=\phi^{(3a)}_{zxy}N_x \delta M_y,
\end{equation}
i.e., an additional staggered component along $z$ is added to the magnetization, when
$\mathbf{N}$ is mainly along $x$, but when a small 
ferromagnetic component is added along $y$ due to Eq.~\eqref{eq_chisp_cof2} as discussed above.

In the following we give an example of how this coexistence of various staggered magnetizations
and cantings affects the symmetry analysis.
Since the 5-th rank axial tensor
\begin{equation}
\phi_{ijklm}^{(5a)}\rightarrow
\langle yzxyy\rangle-\langle xzyxx\rangle
\end{equation}
is allowed by symmetry,
\begin{equation}\label{eq_example_convert_even_to_odd}
\delta\mathcal{O}^{(a,{\rm odd})}_{y}
=\phi_{yzxyy}^{(5a)}
\delta N_{z}
N_{x}
\delta M_{y}
\delta \mathcal{O}^{(a,{\rm even})}_{y}
\end{equation}
predicts a non-staggered axial response along $y$, which is odd in the magnetization,
when $\mathcal{O}^{(a,{\rm even})}_{y}$ is a non-staggered axial response along $y$ even in the magnetization. For example, tensor \#1 predicts
a response of the type  $\mathcal{O}^{(a,{\rm even})}_{y}$, when the polarization vector of the
laser pulse is along [101].
Adding these two contributions,
$\delta\mathcal{O}^{(a,{\rm odd})}_{y}$ and
$\delta \mathcal{O}^{(a,{\rm even})}_{y}$,
we obtain a result that is neither odd nor even in $\mathbf{N}$. 
In fact, the major discrepancy between our \textit{ab-initio} calculations and our simplified
symmetry analysis based on an expansion in $\mathbf{N}$ only is the observation that the
\textit{ab-initio} results are typically neither odd nor even in $\mathbf{N}$ once $\mathbf{N}$ is
tilted away from $z$, while the
symmetry analysis predicts them to be strictly odd or even.
This example explains how this discrepancy may be resolved by considering all aspects of the 
magnetic structure, which generally cannot be captured by a single $\mathbf{N}$.

Alternatively, one may -- of course -- obtain the contribution Eq.~\eqref{eq_example_convert_even_to_odd} also directly from the 6th rank polar tensor
\begin{equation}
\phi^{(6p)}_{ijklmn}
\rightarrow
\langle
yzxyzx
\rangle
+
\langle
xzyxzy
\rangle
\end{equation}
according to the equation
\begin{equation}
    \delta\mathcal{O}^{(a,{\rm odd})}_{y}=
    \phi^{(6p)}_{yzxyzx}\delta N_z N_x \delta M_y E_z E_x^{*}
\end{equation}
in the case where the laser-field polarization is along [101].

The examples above are sufficient to convey the ideas necessary
to extend the simplified symmetry analysis in the general case of $\mathbf{N}$ tilted
away from $z$, while a comprehensive symmetry analysis for this general 
situation is very cumbersome and is beyond the scope of this work.

{\bf Discussion}

One of the key results of our paper are  Eqs.~\eqref{eq:components_relation}$-$\eqref{equation3}  which establish a ``geometric" canted non-relativistic response of the orbital magnetization to incoming linearly-polarized light in rutile altermagnets, where the staggered and non-staggered nature of the components is determined by the orientation of the light polarization with respect to the crystal structure. This canting effect arises as a direct consequence of altermagnetic symmetries, and does not exist in $\mathcal{TS}$-symmetric AFMs. The conclusions of our work are thus generic for all representatives of the rutile altermagnetic class. The degree of canting between the induced moments depends on the relative magnitude of the Cartesian components $-$ which are in principle uncorrelated material-specific quantities $-$ and may even vanish in case one of the components accidentally turns to zero. As the effect is even in $\mathbf{N}$, it is of purely crystal origin, and  may be used to determine the crystal structure of rutile altermagnets given that a direct access to the photo-induced moments for different sense of linearly-polarized light is possible, and $-$ importantly $-$ that is can be disentangled from the SOC-induced contributions.

While serving as prototype altermagnets, rutile materials studied in this work are centrosymmetric. This implies, that in contrast to such $\mathcal{TS}$-symmetric AFMs as Mn$_2$Au~\cite{Reimers_2023} or CuMnAs~\cite{Wadley_2016}, rutile altermagnets do not allow for N\'eel type of torques and spin-orbit torque switching in linear response. In this situation, optical torques, which  from the viewpoint of symmetry  can be directly associated with optically induced second-order transverse spin moments~\cite{freimuth_2016}, present a unique tool for switching the altermagnetic order by light. Such optical torques have been predicted and observed in experiments on ferromagnets~\cite{Choi_2017}, they where shown to serve as a distinct source of  THz emission~\cite{Huisman_2016}, and it was demonstrated that they can trigger optical switching of AFM order in  $\mathcal{TS}$-case~\cite{Ross_2023}. Our non-relativistic calculations predict a staggered ``built-in" orbital response, which, when directly translated into proportional values of spin in a very naive way, implies a staggered spin component and thus a possibility of optical switching of altermagnetism by optical torques. Since the orbital moments do not interact with exchange field directly, SOI is required to translate orbital magnetism into non-equilibrium spin density and thus to exert optical torques.  

Without SOC, as we have seen, the values of non-relativistic orbital moments induced by light are already very large, exceeding in magnitude the orbital moments in the ground state, see Table~\ref{table_3}. Our results demonstrate a drastic impact of SOC on spin and orbital photoresponse. First of all, this manifests in a dramatic increase of induced local spin and orbital moments along the direction of the N\'eel vector. The values that can be induced here under considered  experimental conditions can result in a net magnetization of  about 0.2\,$\mathrm{\mu_B}$, which  already lies in the ballpark of the local staggered spin moments in the ground state of RuO$_2$~\cite{Berlijn_2017}. Among ferromagnets subject to circularly polarized light, this order of magnitude is achievable under normal conditions only for materials containing heavy elements~\cite{freimuth_2016, John_2017}.  Such large values can significantly impact the optically-driven dynamics of the magnetization e.g. when exposed to an external magnetic field in an all-optical-switching type of experiments~\cite{Kimel_2005, Kimel_2019_review}. Detecting this magnetization by MOKE-based techniques~\cite{Wust_2022} may provide a  robust way to determine the direction of the N\'eel vector in altermagnets.  

Manifestly, depending on the direction of N\'eel vector, the SOC-mediated spin and orbital response is not equivalent among the two types of atoms. On the background of large longitudinal induced moments, the transverse moments appear very small, which particularly concerns the case of spin. The same effect of dominant longitudinal response has been also predicted recently for the inverse Faraday effect in $\mathcal{TS}$-symmetric Mn$_2$Au~\cite{Merte_Mn2Au}. However, in the case of considered here altermagnets the  transverse induced spin and orbital moments  possess comparable staggered and non-staggered components at the same time, in analogy to the induced moments along the N\'eel vector. This implies that upon irradiation by  light possibly accompanied by a simultaneous application of a magnetic field, depending on polarization of light and direction of the N\'eel vector, the spins on opposite sublattices will experience radically different optical torques, while the non-equivalent  modifications in the length of the spin moments may give rise to dynamical effects typical of ferrimagnets~\cite{John_2017}. It remains to be explored what complex dynamical regimes can be achieved by light in rutile altermagnets, and what consequences this will have for the staggered magnetization switching, domain wall motion and terahertz radiation by altermagnetism.

{\bf Computational details}
The first-principles electronic structures of RuO$_2$ and CoF$_2$ were calculated by using the full-potential linearized augmented plane wave \texttt{FLEUR} code~\cite{fleurCode}. Exchange and correlation effects were treated with the non-relativistic PBE~\cite{pbe} functional, whereas for relativistic effects the second-variation scheme~\cite{SOC_2nd_var} was employed. The lattice parameters of RuO$_2$ were chosen as $\alpha=4.543$\,a.u and $c=3.140$\,a.u~\cite{Materials_project} and of CoF$_2$ as $\alpha=4.6954$\,a.u and $c=3.1774$\,a.u~\cite{CoF2_Correa}. The muffin-tin radii of Ru and O atoms were set to 2.31\,a.u$^{-1}$ and 1.30\,a.u$^{-1}$, respectively, whereas of Co and F atoms were set to 2.38\,a.u$^{-1}$ and 1.34\,a.u$^{-1}$, respectively. The plane-wave cutoff $K_{\text{max}}$ was set to 4.5\,a.u$^{-1}$ and 4.4\,a.u$^{-1}$ for RuO$_2$ and CoF$_2$, respectively. The semicore $4s$ and $4p$ states of Ru as well as the $3s$ and $3p$ states of Co were treated as local orbitals~\cite{Singh_LOs}. Regarding self-consistent calculations in the Brillouin zone we used a set of 140 $k$-points for RuO$_2$ and a set of 90 $k$-points for CoF$_2$. The AFM ground state configurations were obtained by the use of a Hubbard parameter $U=2.0$\,eV for the $4d$ states of Ru and $U=1.36$\,eV for the $3d$ states of Co, and were consistent with previous first-principles reports on RuO$_2$~\cite{Smejkal_2020} and CoF$_2$~\cite{CoF2_Correa}.
	
Next, we constructed maximally-localized Wannier functions (MLWFs) by employing the Wannier90 code~\cite{Pizzi2020}, where for the initial projections we chose $s$ and $d$ orbitals for Ru and Co atoms, and $p$ orbitals for O and F atoms. We constructed 48 MLWFs out of 72 Bloch functions (BFs) on a 8$\times$8$\times$8 $k$-mesh, with a frozen window of 4.0\,eV above the Fermi energy for RuO$_2$ and 6.5\,eV above the Fermi energy for CoF$_2$.
Our post-processing interpolating calculations of the laser-induced orbital and spin magnetizations were performed on a 128$\times$128$\times$128 $k$-mesh which was sufficient to obtain well-converged results. In all calculations the lifetime broadening $\Gamma$ was set at 25\,meV, the light energy $\hbar\omega$ at 0.25\,meV, the intensity of light at 10\,GW/cm$^2$, and we covered an energy region of $[-2.5, 2.5]$\,eV around the Fermi energy level $\mathcal{E}_F$.

{\bf Acknowledgements}

This work was supported by the Deutsche Forschungsgemeinschaft (DFG, German Research Foundation) $-$ TRR 173/2 $-$ 268565370, TRR
288 – 422213477, Joint Sino-German
Research Projects (Chinese Grant
No. 12061131002 and DFG Grant No.
44880005), and the Sino-German
Mobility Programme (Grant No. 
M-0142). W.F. and Y.Y. are
supported by the NSF of China
(Grant Nos. 12274027, 12321004
12234003). This project has
received funding from the European
Union’s Horizon 2020 research and
innovation programme under the
Marie Skłodowska-Curie grant
agreement No 861300.
We  also gratefully acknowledge the J\"ulich Supercomputing Centre and RWTH Aachen University for providing computational resources under projects  jiff40 and jara0062.


\hbadness=99999 
\bibliography{literature}

\end{document}

%% file: commands.tex
\newcommand{\pgi}{Peter Gr\"unberg Institut,
Forschungszentrum J\"ulich, 52425 J\"ulich, Germany}

\newcommand{\aachen}{Department of Physics, RWTH Aachen University, 52056 Aachen, Germany}

\newcommand{\mainz}{Institute of Physics, Johannes Gutenberg University Mainz, 55099 Mainz, Germany}

\newcommand{\moe}{Centre for Quantum Physics, Key Laboratory of Advanced Optoelectronic Quantum Architecture and Measurement (MOE), School of Physics, Beijing Institute of Technology, Beijing 100081, China}

\newcommand{\beijing}{Beijing Key Lab of Nanophotonics and Ultrafine Optoelectronic Systems, School of Physics, Beijing Institute of Technology, Beijing 100081, China}

\newcommand{\prague}{Institute of Physics, Czech Academy of Sciences, Cukrovarnick\'{a} 10, 162 00 Praha 6, Czech Republic}

\newcommand{\abs}{\mathrm{\abs}}